\documentclass[aps,pra,superscriptaddress,preprint]{revtex4}

\usepackage{graphicx}
\usepackage{graphics}
\usepackage{amssymb}
\usepackage{amsmath}
\usepackage{epsfig}
\usepackage{latexsym}
\usepackage{color}
\usepackage{rotating}
\usepackage{subfigure}

\begin{document}

\title{A classical channel model for gravitational decoherence. }

\author{D. Kafri, J.M.  Taylor}
\affiliation{Joint Quantum Institute,
The University of Maryland }
\author{G. J. Milburn}
\affiliation{Centre for Engineered Quantum Systems, School of Mathematics and Physics, The University of Queensland,  Australia.\\
and Kavli Institute for Theoretical Physics, University of California, Santa Barbara. }

\begin{abstract}
We show that, by treating  the gravitational interaction between two mechanical resonators as a classical measurement channel,  a gravitational decoherence model results that is equivalent to 
 a model first proposed by Diosi.   The resulting decoherence model implies that the classically mediated gravitational interaction between two gravitationally coupled resonators cannot create entanglement. The gravitational decoherence rate ( and the complementary heating rate) is of the order of the gravitationally induced normal mode splitting of the two resonators. 
\end{abstract}
%\pacs{}

\maketitle
\section{Introduction}
The ability to optically cool macroscopic mechanical oscillators close to their ground state, from which highly non-classical superposition states may be prepared,  provides a platform 
 in which to study the interplay between gravitational and quantum physics\cite{Bouwmeester,Pikovski}. The objective is 
to engineer quantum states of mechanical systems in which gravitational effects must be taken into account if we are to account for the dynamics. 
Penrose\cite{Penrose} and also Diosi\cite{Diosi} have  proposed that in such a setting gravity would lead to a new kind of decoherence and, correspondingly, a new source of noise
acting on the quantum degrees of freedom. 

If we had a quantum theory of gravity, the appearance of an additional source of noise would not be remarkable: it would ultimately arise from quantum fluctuations in
the underlying field that mediates the gravitational interaction between quantum mechanical degrees of freedom\cite{Baym}. 
Such effects, it is claimed, are likely to become important at the Planck scale and thus seem unlikely to arise in table-top opto-mechanical experiments for which
the Newtonian description of gravitational interactions would seem to suffice. Surprisingly, the proposals of Penrose and Diosi would indicate that this is not the case and that,
given sufficient quantum control over macroscopic mechanical degrees of freedom, opto-mechanical systems might reveal  gravitational decoherence.  

In this paper we use the recent proposal\cite{Taylor} for classically mediated long range interactions applied to gravitational interactions. In the case of the mutual gravitational interaction of two masses, we model such a classical channel as a  weak continuous measurement of the position of each mass.  The continuous measurement record  is used to control a reciprocal classical force  on each mass via a feedforward control. In essence, the classical measurement record informs each mass of the other's position and applies the corresponding gravitational force. By requiring the effect be completely symmetric in the two masses, we show that the gravitational decoherence rate is completely determined by the gradient of the gravitational force between the two masses. This is equivalent to the gravitational decoherence  models proposed by Penrose and  Diosi.

\section{Combining quantum and gravitational physics. }
\begin{figure}[htbp] %  figure placement: here, top, bottom, or page
   \centering
   \includegraphics[scale=0.5]{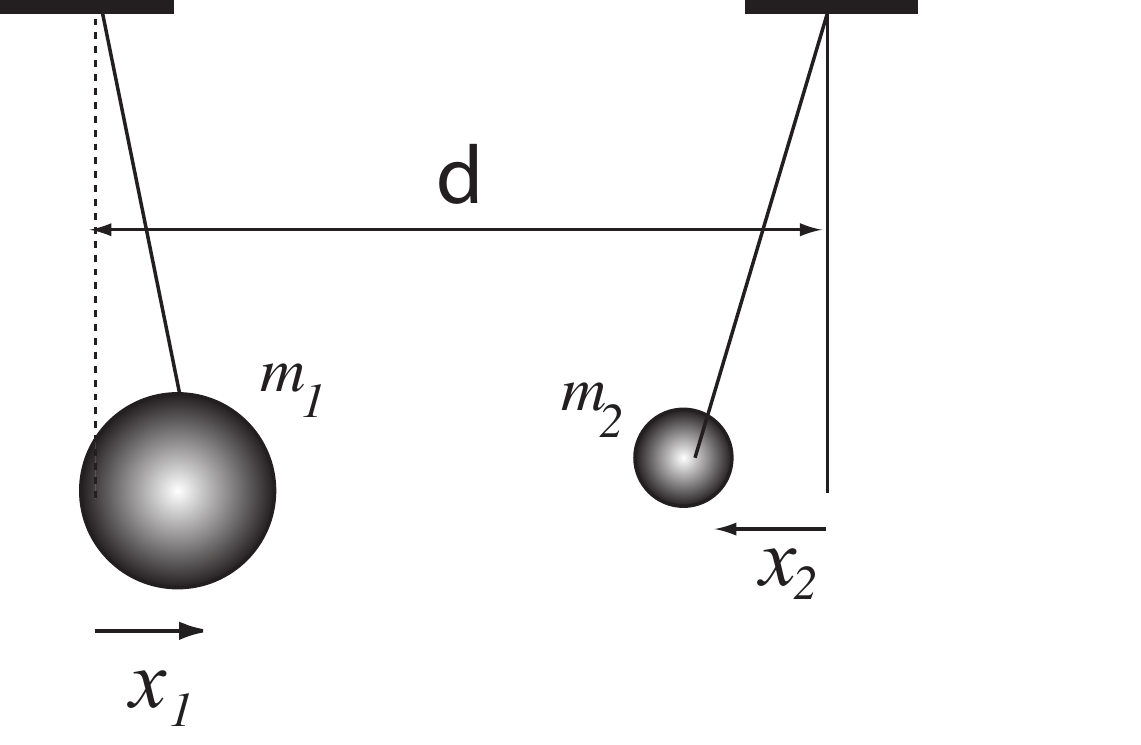} 
   \caption{A gravitationally coupled system of two harmonic oscillators comprising two suspended masses $m_1,m_2$.  }
   \label{fig1}
\end{figure}

Consider two masses,  $m_1,m_2$ freely suspended so as to move (approximately) harmonically along the $x$-axis, Fig. \ref{fig1}. The gravitational interaction
 between the two masses couples these harmonic motions. 
The displacement of mass $m_k$ from equilibrium is denoted $x_k$. The equilibrium positions are determined by the 
gravitational attractions between the two suspended masses. The interaction potential energy between the masses, expanded to second order in the relative displacement, may be written
\begin{equation}
V(x_1,x_2) = V_0-\frac{Gm_1m_2}{d^2}(x_1-x_2) -\frac{Gm_1m_2}{d^3}(x_1-x_2)^2
\end{equation}
The term linear in the displacement represents a constant force between the masses and simply modifies the equilibrium position of the masses  to $\bar{x}_1= Gm_2/(d^2\omega_1^2),\ \ \
\bar{x}_2= -Gm_2/(d^2\omega_2^2)$. We will absorb this into the definition of the displacement coordinates. The quadratic terms proportional to $x_k^2$ can be incorporated into the definition of the harmonic frequency of each mass. The total mechanical Hamiltonian is then given by
\begin{equation}
\label{mech-ham}
H_{qm}=H_0   + K \hat{x}_1\hat{x}_2
\end{equation}
where 
\begin{equation}
H_0=\sum_{k=1}^2\frac{\hat{p}_k^2}{2m_k}+\frac{m_k\Omega_k^2}{2} \hat{x}_k^2
\end{equation}
with 
\begin{equation}
\Omega_k^2=\omega_k^2-K/m_k
\end{equation}
and
\begin{equation}
K=\frac{2Gm_1m_2}{d^3}
\end{equation}
and the usual canonical commutation relations hold $[\hat{x}_k,\hat{p}_j]=i\hbar \delta_{kj}$. 

The model thus reduces to the very well understood case of two quadratically coupled simple harmonic oscillators. The resulting classical and quantum dynamics is then described as two independent simple harmonic oscillators, the normal modes, which are linear combinations of the local co-ordinates $q_+=(x_1+ x_2)/\sqrt{2}$ is the {\em centre-of-mass} mode and $q_-=(x_1-x_2)/\sqrt{2}$ is the {\em breathing} mode with frequencies $\omega_\pm$  given by 
\begin{equation}
\omega_{\pm}^2 =(\Omega_1^2+\Omega_2^2)/2\pm \frac{1}{2}\left [(\Omega_1^2-\Omega_2^2)^2+4K^2/(m_1m_2)\right ]^{1/2}
\end{equation}
In what follows we will consider the symmetric case for which $m_1=m_2=m$ and $\Omega_1=\Omega_2=\Omega$. In that case the normal mode frequencies become
\begin{equation}
\omega_{+} =\omega \ \ \ \\ ; \omega_{-}= \omega\left [1 - \frac{2K}{m\omega^2}\right ]^{1/2}
\end{equation}
In most situations of laboratory relevance,  the gravitational coupling is weak and the difference in frequency between the two normal modes, the normal mode splitting,  can be written
\begin{equation}
\label{mode-split}
\Delta \equiv \omega_+-\omega_-\approx\frac{K}{m\omega}
\end{equation}

\section{Gravity as a classic measurement channel.}
In the holographic view of gravity,  the informational content of the field is carried by {\em classical} information specified on a two-dimensional surface\cite{Bousso}. The idea is motivated by earlier work connecting thermodynamics and space-time curvature in the work of Bekenstein\cite{Bekenstein}, Hawking\cite{Hawking} and Jacobson\cite{Jacobson}. 
Kafri and Taylor\cite{Taylor} have recently proposed a simple way to test if a long range interaction between two particles is mediated by a quantum or a classical channel. They define a quantum channel by introducing an ancillary degree of freedom, a harmonic oscillator. The coherent interactions between two local systems and the channel lead, under appropriate circumstances, to an effective direct non-local interaction between the two local systems.  This kind of process is used in geometric phase gates to simulate non-local interactions between internal states of trapped ions, with the ionic vibrational modes serving as the ancilla\cite{Milburn,Leibfried}. The key of course is the ability to implement controlled entangling operations between the  electronic and vibrational  degrees of freedom of each ion.  A classical channel can then be defined by simply allowing the ancillary oscillator to be continually measured. 

In this paper we will take a different, although equivalent, approach to defining a classical mediated interaction by using methods from quantum stochastic control theory\cite{WisemanMilburn}.

Rather than a direct quantum interaction of the form $\hat{x}_1\hat{x}_2$, we assume the interaction is mediated by a {\em classical channel}. That is, the gravitational centre of mass co-ordinate, $\hat{x}_i$, of each particle is continuously measured and a classical stochastic measurement record, $J_k(t)$, carrying this information acts reciprocally as a classical control force on the other mass. The effect on the dynamics of the systems is to produce a Hamiltonian term of the form,
\begin{equation}
H_{grav} =\chi_1 \frac{dJ_1(t)}{dt} \hat{x}_2 + \chi_2 \frac{dJ_1(t)}{dt} \hat{x}_1 \,.
\end{equation}
 As we will see, the units can be chosen such that  the units of $\chi_k$ are $J m^{-2}$, the same as the units of $K$. 
 
In the case of continuous weak measurements of $\hat{x}_k$  the measurement record obeys a stochastic differential equation of the form\cite{WisemanMilburn}
 \begin{equation}
d J_k(t) =\langle \hat{x}_k\rangle_c dt+\sqrt{\frac{\hbar}{2\Gamma_k}}dW_k(t)
 \end{equation}
 where $\Gamma_k$ is a constant that determines the rate at which information is gained by the measurement, while $dW_{1,2}$ are independent, real valued Wiener increments.  The units of $\Gamma_k/\hbar$ are $m^{-2}s^{-1}$. The average $\langle \hat{x}_k\rangle_c$ is a {\em conditional} quantum mechanical average conditioned on the entire history of measurement records up to time $t$. 

 The conditional quantum dynamics of the opto-mechanical system is given by the stochastic master equation
 \begin{equation}
 \label{X-conditional}
  d\rho_c =  -\frac{i}{\hbar}[H_{c},\rho_c]dt-\sum_{k=1}^2\frac{\Gamma_k}{2\hbar}[\hat{x}_k,[\hat{x}_k,\rho_c]]dt+\sqrt{\frac{\Gamma_k}{\hbar}}dW_k(t){\cal H}[\hat{x}_k]\rho_c
  \end{equation}
  where the classical control Hamiltonian is defined by
 \begin{equation}
 \label{holographic-ham}
H_{c}= H_0+H_{grav}
\end{equation}
and the conditioning super-operator, ${\cal H}$ is defined by
\begin{equation}
{\cal H}[\hat{X}]\rho =\hat{X}\rho+\rho\hat{X}-{\rm Tr}\left (\hat{X}\rho+\rho\hat{X}\right )
\end{equation}
The appearance of the conditional average, ${\rm Tr}\left (\hat{x}_k\rho+\rho\hat{x}_k\right )$ in Eq.(\ref{X-conditional}) is very similar to the theory of \cite{Halliwell} and also \cite{Yanbei} although in the latter case the stochastic term is missing and the interpretation is thus very different.

The form of Eq. (\ref{holographic-ham}) defines a direct feedback model of the kind considered in \cite{WisemanMilburn}. Using the results there we find that the corresponding unconditional dynamics is given by 
 \begin{eqnarray}
 \label{full-me}
  \frac{d\rho}{dt}  & =  &  -\frac{i}{\hbar}[H_0,\rho]-\frac{i}{2\hbar}\left (\chi_2[\hat{x}_1, \hat{x}_2\rho+\rho\hat{x}_2]+\chi_1[\hat{x}_2, \hat{x}_1\rho+\rho\hat{x}_1]\right )\\\nonumber
  & & -\frac{\Gamma}{2\hbar}\sum_{k=1}^2[\hat{x}_k,[\hat{x}_k,\rho]]-\frac{1}{8\hbar\Gamma}\sum_{k=1}^2\chi_k^2[\hat{x}_k,[\hat{x}_k,\rho]]
\end{eqnarray}
The second term is the systematic effect of the control protocol. It is easy to see that if we fix $\chi_1=\chi_2=K$ this reduces to the standard Hamiltonian interaction term given in Eq.(\ref{mech-ham}).  The noise added by measurement and feedback is a minimum at $\Gamma=\chi/2$ linking the decoherence rate  due to the continuous measurement to the scale of the gravitational interaction   as 
\begin{equation}
\Gamma=K/2
\end{equation}
so that the rate at which classical information is transmitted by the classical channel is determined entirely by the gradient of the gravitational field.
The resulting  unconditional dynamics is
  \begin{equation}
  \label{grav-me}
  \frac{d\rho}{dt} =  -\frac{i}{\hbar}[H_0,\rho]-\frac{i}{\hbar}K[\hat{x}_1\hat{x}_2, \rho] -\frac{K}{2\hbar}\sum_{k=1}^2[\hat{x}_k,[\hat{x}_k,\rho]]
  \end{equation}
 This is consistent with Diosi's model \cite{Diosi} which gives the same decoherence rate as obtained here under similar approximations \cite{Diosi-2007}.  
The form of Eq.(\ref{grav-me}) can be generalized to exactly match the one in reference \cite{Taylor} if we include non-cross terms in the feedback. I.e., in equation (9), we could have terms proportional to $(d J_1 /dt) x_1$ and   $(d J_2 /dt) x_2$.
 
 There are similarities between Diosi's approach and the measurement mediated approach described here. Both require  that the gravitational interaction between 
the two degrees of freedom be replaced with a noisy interaction. In the measurement based approach this is incorporated in a way which necessarily preserves positivity as the noise arises from a `hidden' position measurement of the gravitational centre of has co-ordinate.  In Diosi's approach we need to explicitly constrain the noise to preserve positivity. More recently Diosi\cite{Diosi2014} has pointed out that a system subject to weak continuous measurement is a natural example of a quantum-classical hybrid dynamics.   

Using a recent result of Kafri and Taylor\cite{Taylor}, we can show that, in the case when the two systems are Gaussian, the master equation Eq. (\ref{grav-me}) can never entangle them. This is also true for Eq. (\ref{full-me}), assuming $\chi_1 = \chi_2$. Conversely, if the decoherence effect were any weaker, we could entangle the ground state.  Further, the gravitational decoherence in the dynamics is minimal in the sense that, if it were any smaller, evolution under Eq. (\ref{grav-me}) would immediately entangle the ground state of the (uncoupled) Hamiltonian, $H_0$.

To see this we write the Eq.(\ref{grav-me}) in terms of the dimensionless operators $\tilde{x}_k=x_k\left (m\omega/\hbar\right )^{-1/2}$,
\begin{equation}
\frac{d\rho}{dt} = -\frac{i}{\hbar} [H,\rho]-ig[\tilde{x}_1\tilde{x}_2,\rho]-\frac{1}{4}\sum_{k=1}^2Y_{ij}[\tilde{x}_i,[\tilde{x}_j,\rho]]
\end{equation}
where $g=\frac{K}{m\omega}$ measures the strength of the gravitational interaction and the matrix $Y_{ij}=\left (\frac{2K}{m\omega}\right )\delta_{ij}$ is the decoherence matrix. Using the result from \cite{Taylor}, we note that entanglement is never generated if and only if the matrix $Y-2ig\sigma$ has no negative eigenvalues, where $\sigma$ is the $2\times 2$ symplectic matrix 
\begin{equation}
\sigma=\left ( \begin{array}{cc}
     0 & 1 \\
      -1 & 0\end{array}\right )
      \end{equation}
 Noting that $Y-2ig\sigma$ has eigenvalues $0$ and $4g$  we see that a slightly less noisy matrix $Y_{ij}-\epsilon\delta_{ij}$ produces entanglement for any positive $\epsilon$. 

%To see this we write the Eq.(\ref{grav-me}) in terms of the dimensionless operators $\tilde{x}_k=x_k\left (m\omega/\hbar\right )^{-1/2}$
%\begin{equation}
%\frac{d\rho}{dt} = -\frac{i}{\hbar} [H,\rho]-ig[\tilde{x}_1\tilde{x}_2,\rho]-\frac{1}{4}\sum_{k=1}^2Y_{ij}[\tilde{x}_i,[\tilde{x}_j,\rho]]
%\end{equation}
%where $g=\frac{K}{m\omega}$ measures the strength of the gravitational interaction and the matrix $Y_{ij}=\left (\frac{2K}{m\omega}\right )\delta_{ij}$ is the decoherence matrix. Using the result from \cite{Taylor} we then see that entanglement is never generated if and only if the matrix $Y-2ig\sigma$ has no negative eigenvalues where $\sigma$ is the $2\times 2$ symplectic matrix 
%\begin{equation}
%\sigma=\left ( \begin{array}{cc}
%     0 & 1 \\
%      -1 & 0\end{array}\right )
%      \end{equation}
% Noting that $Y-2ig\sigma$ has eigenvalues $0$ and $4g$  we see that a slightly less noisy matrix $Y_{ij}-\epsilon\delta_{ij}$ produces entanglement for any positive $\epsilon$.  

\section{An experimental test of gravitational decoherence.}
We now consider the prospects for an experimental observation of the model proposed here. 
For simplicity we will assume that the two mechanical resonators have the same mass ($m_1=m_2=m$) and frequency ($\omega_1=\omega_2=\omega$).
The last term in Eq. (\ref{grav-me}) is responsible for two complementary effects:  it drives a diffusion process in momentum of each of the
oscillators at the rate $\hbar K$, which we will call the gravitational heating rate
\begin{equation}
D_{grav} =\hbar K
\end{equation}
The momentum diffusion leads to heating of the mechanical resonators. It is convenient to define this in terms of the rate of change of the 
phonon number; the average mechanical energy divided by $\hbar\omega$. The heating rate is then given by
\begin{equation}
R_{grav}=\frac{K}{2m\omega}
\end{equation}

The double commutator term also leads to the decay of off-diagonal coherence in the position basis of each mechanical resonator,
\begin{equation}
\frac{d\langle x'_k|\rho|x_k\rangle}{dt}=(\ldots)-\frac{K}{2\hbar}(x'_k-x_k)^2
\end{equation}
This shows that the rate of decay of coherence is more rapid the greater the separation of the superposed states. 
We can use the natural length scale proceeded by the zero-point  position fluctuations in the ground state of each resonator to 
rewrite the decoherence rate as
\begin{equation}
\Lambda_{grav}=\frac{K}{2\hbar}\Delta x_0^2=\frac{K}{4m\omega}
\end{equation}
Thus the gravitational decoherence rate for position, in natural units, is one half the gravitational heating rate.

These rates can equivalently be expressed in terms of the normal mode splitting when the gravitational interaction is weak, Eq. (\ref{mode-split}). 
\begin{eqnarray}
R_{grav}  & = & \frac{\Delta}{2} \\
\Lambda_{grav} & = & \frac{\Delta}{4}
\end{eqnarray}
We thus see that the key parameter responsible for gravitational decoherence is of the order of the normal mode 
splitting between the two mechanical resonators due to their gravitational coupling. This has significant consequences for 
observation. 

In order to see  gravitational decoherence in this model, we need to arrange for the normal mode splitting to be as large as possible. Writing 
this in terms of the Newton constant, we see that
\begin{equation}
\Delta = \frac{Gm}{\omega d^3}
\end{equation}
In the case of two spheres of radius $r$, this may be written in terms of the density of the material as 
\begin{equation}
\Delta =\frac{4\pi G\rho}{\omega} \left (\frac{r}{d}\right )^3
\end{equation}
 As $d<2r$, this quantity is bounded
 \begin{equation}
\Delta \leq \frac{\pi G\rho}{6 \omega} 
\end{equation}
We need to use a material with a large density and a mechanical frequency as small as possible. For example, for depleted uranium spheres and a mechanical frequency of one Hertz, we find that $\Delta \sim10^{-7} {\rm s}^{-1}$, a value so small that a terrestrial experiment would be challenging. 

In a realistic experiment with low frequency mechanical resonators of the kind considered here, thermal noise and frictional damping will be unavoidable. We can estimate the relative size of these effects using the quantum Brownian motion master equation\cite{WisemanMilburn},
\begin{equation}
\left . \frac{d\rho}{dt}\right |_{diss} =  \sum_{j=1}^2-i\gamma_k[\hat{x}_j,\{\hat{p}_j,\rho\}]-2\gamma_j k_BT m_j[\hat{x}_j,[\hat{x}_j,\rho]]
\end{equation}
where $\gamma_k$ is the dissipation rate for each of the mechanical resonators assumed to be interacting with a common thermal environment at temperature $T$. If we compare the form of 
thermal noise in this equation to the form of gravitational decoherence, for the symmetric case,  we see that we can assign an effective temperature to the gravitational decoherence rate given by
\begin{equation}
T_{grav} =  \frac{\hbar K}{2m\gamma k_B}
\end{equation}
If we write this in terms of the quality factor, $Q$, for the mechanical resonators, it gives an effective thermal energy scale of 
 \begin{equation}
k_BT_{grav} = \hbar Q \Delta
\end{equation}
In the example discussed in the previous paragraph for  the relatively high value of $Q=10^9$ we find that $T_{grav}\sim 10^{-9}$K.   One would need an ambient temperature less than this to clearly distinguish gravitational decoherence from environmental effects.  It is difficult to see how this might be achieved even using modern opto-mechanical laser cooling of mechanical resonators. Possibly gravitationally coupled Bose-Einstein condensates of atomic gases could reach this regime.  

\section{Conclusion}
In this paper we have presented a model in which the force of gravity is mediated by a purely classical channel. The channel is defined by considering a continuous weak measurement of the position of each of two masses and feeding forward the classical stochastic record of measurement results to induce the right gravitational force between the masses. As the weak measurement model is entirely consistent with quantum mechanics, exactly the right amount of noise is introduced to ensure that the resulting master equation, when averaged over all measurement records, is positivity preserving. Using a minimal symmetric argument, we find that the model is equivalent to a gravitational decoherence  model first proposed by Diosi. As the systematic effect of the gravitational interaction also fixes the size of the noise, minimising the noise leads to a model  with no free parameters. 

Our model is a specific example of a general theory of a classically mediated (i.e. non-entangling) force law considered  by Kafri and Taylor \cite{Taylor}. Using a result in that paper we find that a classically mediated gravitational channel based on continuous weak measurement  can never entangle Gaussian systems. In an experimental setting of two identical gravitationally coupled resonators, this result is manifest as a direct scaling between the normal mode splitting induced by the gravitational force and the gravitational decoherence rate.  An experimental test using two gravitationally coupled opto-mechanical resonators would be difficult, but not impossible, with current technology. 

\acknowledgements
GJM wishes to acknowledge useful discussions with Tim Ralph, Casey Myers and Nathan McMahon. This work was partly supported by the Australian Research Council grant CE110001013.


\begin{thebibliography}{99}
\bibitem{Bouwmeester}W. Marshall, C. Simon, R. Penrose, and D. Bouwmeester, Phys. Rev. Letts., {\bf  91},  130401 (2003).
\bibitem{Pikovski}I. Pikovski, M. R. Vanner, M. Aspelmeyer, M. S. Kim and C. Brukner, Nature Physics,  {\bf 8}, 393, (2012). 
\bibitem{Penrose}R. Penrose, Gen. Rel. Grav. {\bf 28} 581 (1996).
\bibitem{Diosi}L. Diosi, {\em The gravity-related decoherence master equation from hybrid dynamics}, J. Phys.: Conf. Ser. {\bf 306}, 012006 (2011). 
\bibitem{Baym}G Baym and T. Ozawa, PNAS, {\bf 106}, 3035 (2009). 
\bibitem{Taylor}D. Kafri and J. M. Taylor, {\em A noise inequality for classical forces},  arXiv:1311.4558v1 (2013) 
%\bibitem{Karolyhazy}F.Karolyhazy, Nuovo Cimento A42  390 (1966). 
%\bibitem{Soni}Vikram Soni Pramana 59:375-384,(2002): see also arXiv:gr-qc/0009065. 
\bibitem{Yanbei}Huan Yang, Haixing Miao,  Da-shin Lee, Bassam Helou, and Yanbei Chen, arXiv:1210.0457v2 (2012). 
%\bibitem{Moller}C. Moller, Les Theories Relativistes de la Gravitation Collo- ques Internationaux CNRX 91 ed A Lichnerowicz and M-A Tonnelat (Paris: CNRS) 1962
%\bibitem{Rosenfield}L. Rosenfeld, Nucl. Phys. 40 353 (1963).
%\bibitem{Carlip}S. Carlip, Class. Quantum Grav. 25 154010 (2008).
\bibitem{Halliwell}J. J. Halliwell, Phys Rev D {\bf 57}, 2337 (1998)
\bibitem{Milburn}G.J. Milburn, S. Schneider and D.F.V. James, Fortschritte der Physik,  48, 801 (2000). 
\bibitem{Leibfried}D. Leibfried, B. DeMarco, V. Meyer, D. Lucas, M. Barrett, J. Britton, W. M. Itano, B. Jelenkovic, C. Langer, T. Rosenband and D. J. Wineland, Nature {\bf 422}, 412 ( 2003)
\bibitem{Bousso}R. Bousso, Rev. Mod. Phys. {\bf 74}, 825 (2002). 
\bibitem{Bekenstein}J. D. Bekenstein, Phys. Rev. D. {\bf 23}, 287 (1981).
\bibitem{Hawking}S. W. Hawking, Comm. Math. Phys. {\bf 43}, 199 (1975).
\bibitem{Jacobson}T. Jacobson, Phys.Rev.Lett. {\bf 75}, 1260 (1995).
%\bibitem{Verlinde}E. P. Verlinde, arXiv:1001.0785(2010).
\bibitem{WM}G. J. Milburn and M. J. Woolley {\em An introduction to quantum optomechanics}, Acta Physica Slovaca, {\bf 61}, no. 5 (2012). 
\bibitem{WisemanMilburn} H. M. Wiseman and G. J. Milburn {\em Quantum Measurement and Control}, (Cambridge University Press, Cambridge, 2010).
\bibitem{SM}A.J.Scott and G.J.Milburn, Phys. Rev. A {\bf 63}, 042101 (2001).
\bibitem{Milburn-RS}G. J. Milburn, G. J. Milburn, Proc. Roy Soc. A, 370, 4469 (2012). 
%\bibitem{Retzker}A Retzker, J Cirac, B Reznik, Phys. Rev. Letts., {\bf 94}  050504 (2005).
\bibitem{Diosi-2007}L. Diosi, J. Phys. A: Math. Theor. {\bf 40},  2989 (2007))
\bibitem{Diosi2014}L.Diosi, {\em Hybrid Quantum-Classical Master Equations}, arXiv:1401.0476 (2014). 

\end{thebibliography}
\end{document}